\documentclass[12pt,a4paper]{article}

\usepackage[english]{babel}

\usepackage[letterpaper,top=2cm,bottom=2cm,left=3cm,right=3cm,marginparwidth=1.75cm]{geometry}

\usepackage{amsmath}
\usepackage{graphicx}
\usepackage[colorlinks=true, allcolors=blue]{hyperref}

\usepackage{listings} 
\usepackage{array} 
\usepackage{xcolor}
\usepackage{adjustbox}

\definecolor{codegray}{rgb}{0.95,0.95,0.95}
\definecolor{commentgray}{rgb}{0.4,0.4,0.4}
\definecolor{keywordblue}{rgb}{0.2,0.2,0.7}
\definecolor{stringred}{rgb}{0.8,0.1,0.1}
\definecolor{linenumbergray}{rgb}{0.5,0.5,0.5}
\definecolor{framegray}{rgb}{0.75,0.75,0.75}

\lstdefinelanguage{Solidity}{
  morekeywords={
    contract,function,modifier,event,enum,struct,if,else,while,for,import,return,
    mapping,address,bool,string,public,private,internal,external,view,pure,storage,memory,new,
    require,assert,revert,emit,calldata,override,virtual,constructor
  },
  sensitive=true,
  morecomment=[l]{//},
  morecomment=[s]{/*}{*/},
  morestring=[b]",
}

\title{DoS Attacks and Defense Technologies in Blockchain Systems: A Hierarchical Analysis}

\author{CHUNYI ZHANG\footnotemark[1], FENGJIAO DOU\footnotemark[1], XIAOQI LI\footnotemark[2]}

\date{}

\begin{document}
\maketitle

\renewcommand{\thefootnote}{*}
\footnotetext[1]{Chunyi Zhang and Fengjiao Dou contributed equally to this work.}
\footnotetext[2]{Authors’ Contact Information: Chunyi Zhang, Hainan University, Haikou, Hainan, China; Fengjiao Dou, Hainan University, Haikou, Hainan, China; Xiaoqi Li, csxqli@ieee.org, Hainan University, Haikou, Hainan, China.}

\begin{abstract}
Blockchain technology is widely used in various fields due to its ability to provide decentralization and trustless security. This is a fundamental understanding held by many advocates, but it is misunderstood, leading participants to fail to recognize the limitations of the security that blockchain can provide. Among all current network attacks, Denial of Service (DoS) attacks pose significant threats due to their ease of execution and destructive potential. This paper, based on the blockchain architecture hierarchy, categorizes and organizes existing DoS attacks, with a focus on explaining the principles and methods of contract layer and consensus layer DoS attacks. Furthermore, this paper comprehensively analyzes and compares commonly used detection methods and defense technologies, which will contribute to strengthening the security and stability of blockchain systems and promoting further innovation and application of blockchain systems.
\end{abstract}

\section{Introduction}
Blockchain is a new application model of computer technology, such as distributed data storage, peer-to-peer transmission, consensus mechanism, and encryption algorithm. Essentially, it is a decentralized shared database that stores data or information that is unforgeable, traceable, open, transparent, and collectively maintained.

Blockchain technology has made remarkable progress in the international arena and is widely used in other fields such as smart contracts, supply chain, finance, healthcare, and government \cite{guo2022survey}. This trend of wide application has made it a key technology support in many fields. For example, in the field of smart contracts, blockchain technology brings increased efficiency, transparency, and security. Within the domain of supply chains, it has been shown to promote transparency and improve traceability~\cite{arachchige2024analysis}. Moreover, in the field of finance, blockchain has been found to improve cross-border payments, asset management, and transaction settlement. Finally, in the area of healthcare, it provides a secure framework for managing and sharing health data \cite{li2024stateguard}.

The development and wide application of blockchain technology have also raised concerns about its stability and security. In recent years, there have been several DoS attacks on blockchain systems that have caused major consequences, including system failure and property damage, among which the more influential ones are \cite{konig2020risks}:

(1) Bitcoin Network Transaction Fog Attack. In 2014, the attacker generated a substantial number of small transactions, each comprising numerous inputs and outputs \cite{zhong2023tackling}. This led to the accumulation of a large amount of transaction data. These substantial transactions effectively occupied the available space within the block, thus impeding the processing of routine transactions. This phenomenon is referred to as transaction fog, which occurs when the network becomes opaque, hindering the visibility of the actual transactions~\cite{zhang2024denial}.

(2) Bitcoin Network Stress Test. In 2015, the incident was initially conceived as a stress test to assess the processing capacity of the Bitcoin network. However, the magnitude of the test precipitated a Distributed Denial of Service (DDoS) attack \cite{bu2025smartbugbert}. During the test, the volume of transactions on the network increased exponentially, resulting in a substantial number of transactions that could not be processed promptly. This, in turn, affected the normal transactions of Bitcoin.

(3) DDoS attack on Ether. In 2016, attackers took advantage of vulnerabilities in smart contracts, resulting in the creation of a substantial number of transactions that used a significant amount of computational resources \cite{li2021hybrid}. This led to a substantial reduction in the processing capacity of the Ether network.

(4) KotET Incident. In 2016, the deployment of smart contracts for the KotET game was completed, in which the player sends the contract some ETH (Ethereum) to obtain the throne \cite{li2017discovering}. The Ponzi trap occurs because as the number of kings increases, the cost of becoming a king also increases.

(5) DDoS attack on the IOTA network. IOTA is a cryptocurrency that uses Directed Acyclic Graph (DAG) technology, which is different from the traditional blockchain structure. In 2017, IOTA suffered a DDoS attack in which attackers overloaded IOTA's network nodes by sending a large number of transaction requests in succession, preventing them from processing normal transactions \cite{bu2025enhancing}.

(6) Microsoft suffered a 2.4Tbps DDoS attack. In 2021, Microsoft acknowledged that it had been subjected to a 2.4Tbps DDoS attack in August, representing a 140 percent increase compared to the previous highest recorded attack bandwidth by Microsoft in 2020 \cite{liu2024gastrace}.

These examples demonstrate that, despite the decentralized nature of the blockchain, which renders it more resistant to DoS attacks than traditional centralized networks, it is not entirely impervious to vulnerabilities. While it cannot fully incapacitate the network as can traditional DDoS attacks, it can result in the degradation of network service quality \cite{wu2025exploring}. Consequently, for blockchain designers and operators, although blockchain networks generally possess certain mechanisms (e.g., transaction fees, computational difficulty, etc.) to impede such attacks, the prevention and response to DoS attacks remains an issue that necessitates continuous attention \cite{chen2022survey}.

The advent of blockchain technology and the subsequent revelation of its vulnerabilities have prompted the academic community to undertake comprehensive research endeavors concerning its security \cite{zeng2020survey}.
In 2018, Cai Z, Du C, and Gan Y provided a concise overview of classification, architecture, and key blockchain technologies~\cite{cheng2022cooperative}. They conducted a preliminary analysis of key management, access control mechanisms, defense mechanisms against DDoS attacks, fragmentation leak prevention mechanisms, and future blockchain development trends. In addition, they predicted the security mechanisms of the blockchain. This analysis was intended to facilitate further research on the blockchain and its security \cite{cai2018research}. Subsequently, Gu Xin and Xu Shuzhen et al. expounded on the archetypal blockchain technology architecture, meticulously analyzing the security of blockchain technology from the perspectives of block data structure, hash algorithms, digital signatures, and smart contracts. In subsequent years, a number of researchers and scholars have initiated the discussion of issues pertaining to the security and privacy of Bitcoin and the blockchain \cite{zaghloul2020bitcoin}. In 2021, Tian et al. conducted a comprehensive analysis of the security and privacy of Bitcoin. Their analysis encompassed various aspects, including hierarchical classification and attack attribution analysis, among others. This study aimed to analyze the system architecture, attack principles, and defense strategies of existing security problems on blockchain by examining the dimensions of Layer Classification and Attack Correlation Analysis \cite{saad2020exploring}. In 2022, Yang et al. presented a comprehensive analysis of 14 prevalent security vulnerabilities in smart contracts and summarized methods of vulnerability-related security prevention~\cite{singh2024blockchain}.  They analyzed four aspects of data storage security, data privacy security, data access security, and data sharing security, which provided some reference for future research work of related people in the field of blockchain security \cite{guru2021approaches}.

In recent years, many scholars have conducted detailed overviews from the development history of blockchain technology, quantitative comparison of consensus algorithms, cryptographic details in public-key cryptography, zero-knowledge proofs used in blockchain, hash functions, comprehensive lists of blockchain applications, etc., to assess blockchain security from the perspective of risk analysis and come up with a more comprehensive category of blockchain security risks \cite{anand2022blockchain}.
The focus of academic research is predominantly on the analysis of algorithms, protocols, implementations, and the utilization of security measures. However, there is a paucity of research on man-made attacks and external threats that, given the ease with which they can be implemented, can inflict significant damage through DoS attacks. Among the various types of network attacks, these external attacks pose the greatest threat.

This study begins with the layered structure of blockchain technology, analyzes existing DoS attack cases, and investigates typical attack vulnerability principles and the corresponding defense technologies at each layer. First, the current state of academic research on attacks and defenses targeting blockchain systems is described, with a focus on the foundational theories relevant to this paper. Second, starting from the layered structure, each layer of attacks is analyzed with examples, and classic attack methods are reproduced. Then, the paper compares the advantages and disadvantages of common detection and defense schemes across the four layers, with a focus on contract vulnerability detection. Finally, this paper identifies current challenges and difficulties, summarizes research findings, and outlines directions for future research improvements.

The main contributions of this study are:
\begin{itemize}
\item \textbf{Layered attack classification framework:} We propose a systematic DoS attack classification method based on the seven-layer architecture of the blockchain, defining potential threat models for each layer.
\item \textbf{Defense technology comparative assessment:} We propose the corresponding defense solutions for attacks at various levels and systematically compare their advantages and disadvantages.
\item \textbf{Attack reproduction and vulnerability detection:} We simulate and reproduce a DoS attack on smart contracts and use Mythril to detect timestamp dependency vulnerabilities, providing reproducible vulnerability exploitation paths and repair solutions.
\end{itemize}

This study aims to fill the gap in the research on DoS attacks in the blockchain field and promote further innovation and application of blockchain systems.

\section{Background}
\subsection{Blockchain}
The fundamental principle of blockchain technology is decentralization, which is the process of distributing data across multiple nodes in a network, thereby ensuring its reliability and security. This decentralized approach offers numerous advantages, including immutability, transparency, traceability, and retroactivity \cite{li2024defitail}.

\subsubsection{Block Structure}
Blockchain is a distributed database in which data is linked in the form of blocks. Each block contains a certain amount of data and metadata associated with it. Blockchain technology is characterized by a block-to-chain structure that facilitates decentralized, tamper-proof, transparent, and trustworthy data storage and transmission \cite{liu2025sok}.

\begin{figure}[t]
\centering
\includegraphics[width=0.68\linewidth]{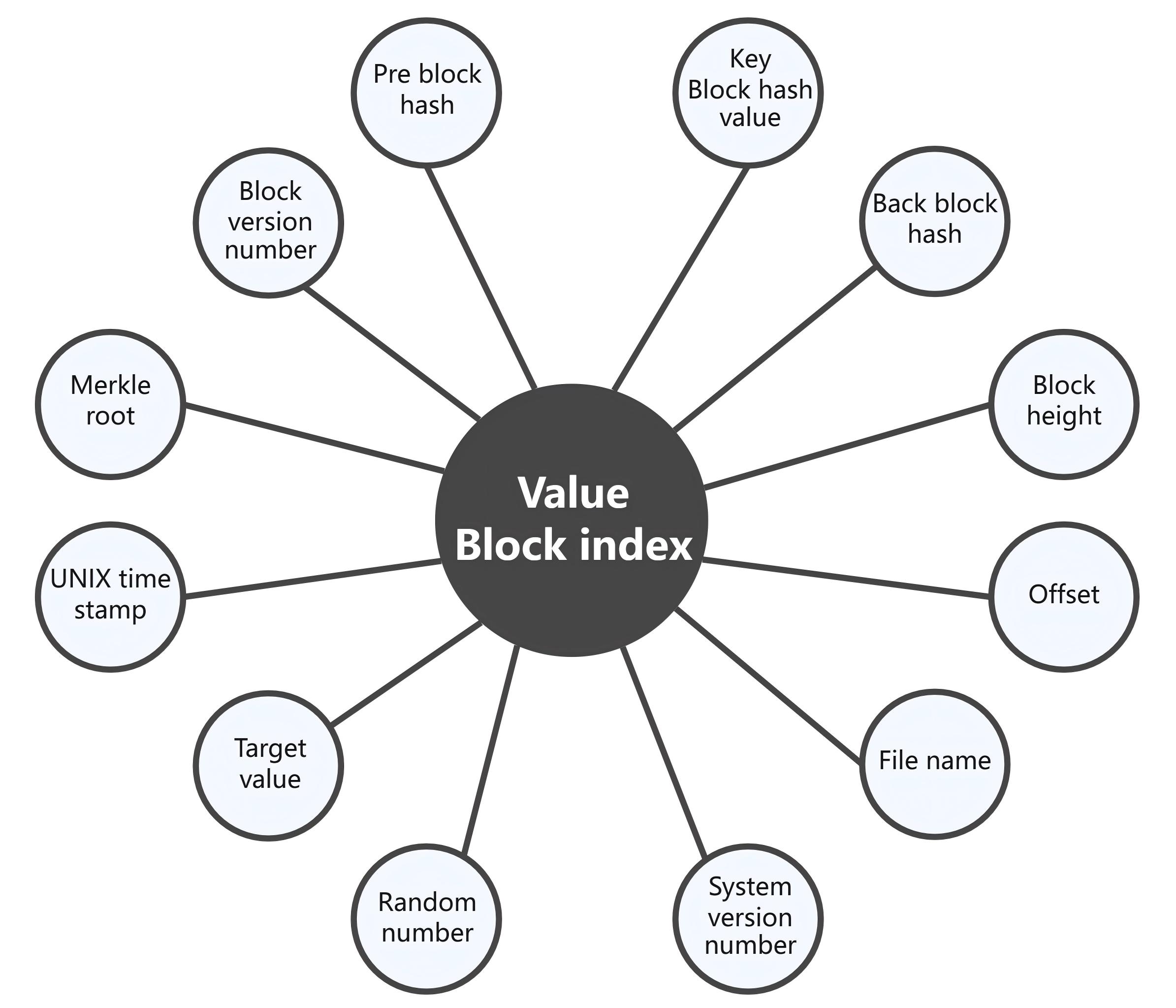}
\caption{\label{fig:Block Index}Block Index}
\end{figure}

\textbf{Block.} The blockchain is composed of blocks as the basic unit and contains metadata. Metadata is extremely important data in a block, including timestamps, hashes of previous blocks, random numbers (Nonce), etc~\cite{chaganti2023survey}. Each block usually contains data on one or more transactions, which can be digital currency transfers, smart contract executions, etc. 
The block index is shown in Figure \ref{fig:Block Index}.

\textbf{Hash.} The hash algorithm is irreversible, thereby excluding the possibility of recovering the original data from the encrypted version~\cite{jiang2024credible}. According to its underlying principle, it can be seen that only slightly changing the data in the block will lead to changes in the hash value. Therefore, the transaction data in the block is encrypted and verified by hashing, which ensures its integrity and security.

\begin{figure}[t]
\centering
\includegraphics[width=0.95\linewidth]{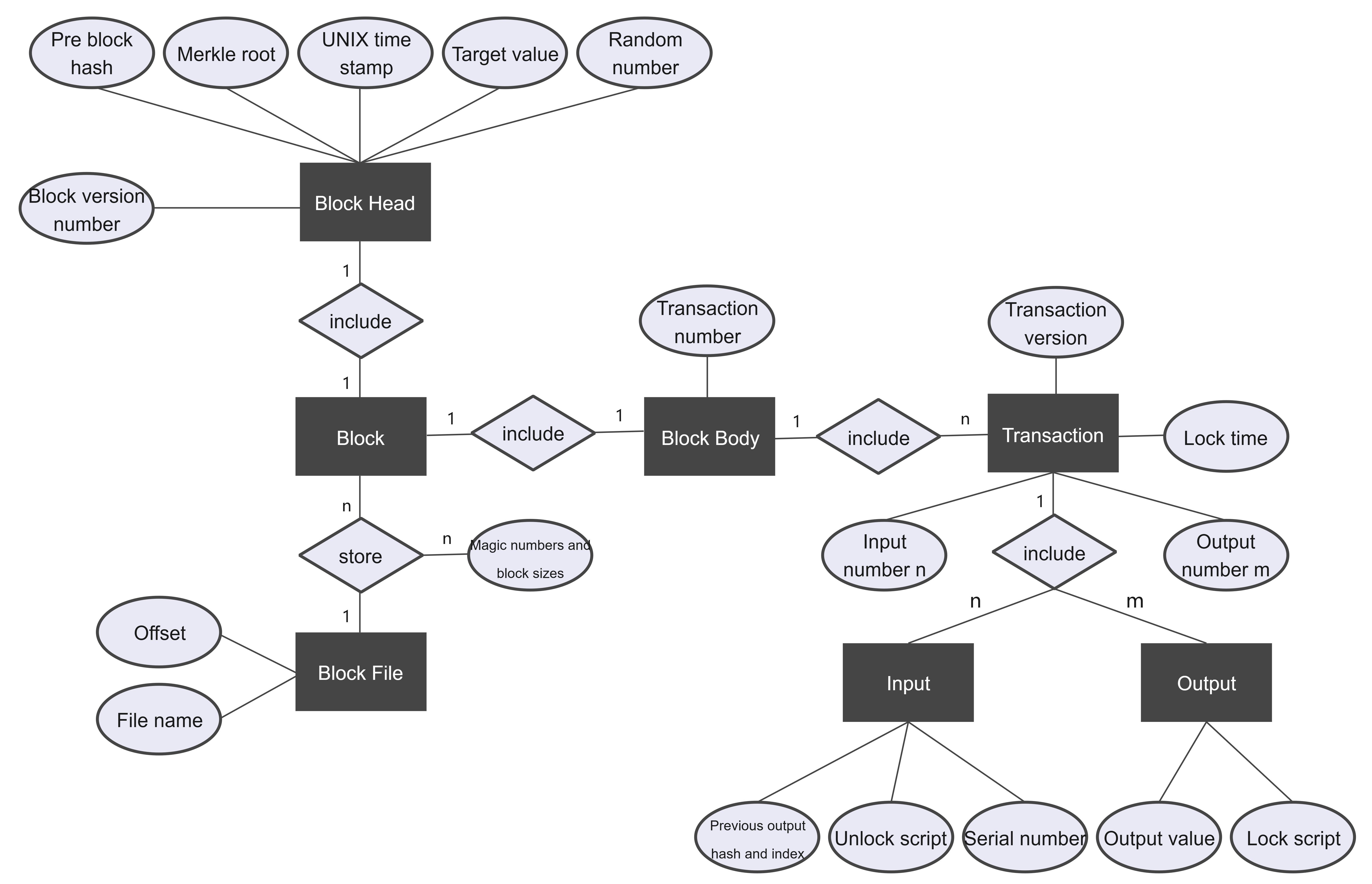}
\caption{\label{fig:From Block to Chain}From Block to Chain}
\end{figure}

\textbf{Chain.} The blockchain is composed of a series of sequentially linked blocks. Each block contains data, including the hash value of the previous block~\cite{alam2023use}. This results in the data forming a continuous chain on the blockchain. Each node has a complete copy of the blockchain. The blockchain structure from block to chain is shown in Figure \ref{fig:From Block to Chain}.

\textbf{Consensus Mechanism.} The consensus mechanisms ensure that all nodes in the network agree to a specific action, thus reaching a consensus~\cite{saveetha2024integrated}. These mechanisms also determine the validity of specific branches, including Proof of Work (PoW), Proof of Stake (PoS), and Proof of Equity Shares (PoA).

\subsubsection{Blockchain Architecture}
\begin{figure}[t]
\centering
\includegraphics[width=0.8\linewidth]{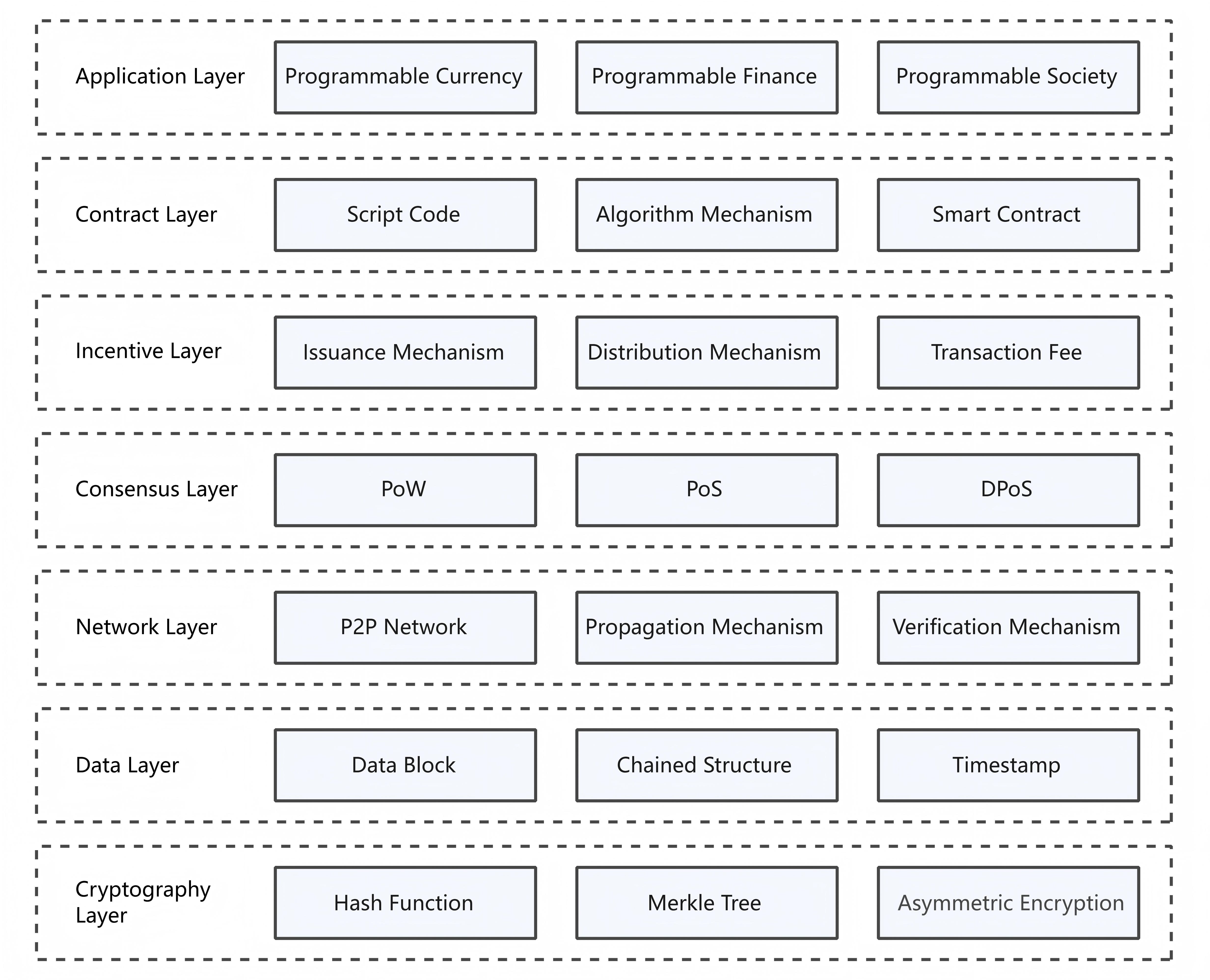}
\caption{\label{fig:Blockchain Architecture}Blockchain Architecture}
\end{figure}

In blockchain architecture, the blockchain is divided into seven layers: cryptographic layer, data layer, network layer, consensus layer, incentive layer, contract layer, and application layer \cite{li2024cobra}. In the seven-layer architecture of blockchain, each layer has different functions and roles. The diagram of the blockchain structure is shown in Figure \ref{fig:Blockchain Architecture}.

\textbf{Cryptography Layer.} The cryptography layer, situated at the foundation of the blockchain, employs a range of cryptographic theories, including encryption algorithms, digital signatures, and hash functions, to ensure the security and confidentiality of data and communications within the blockchain system~\cite{wang2023optimal}. The implementation of encryption technology ensures the confidentiality, integrity, and protection of transaction data, protecting against potential threats such as tampering and theft.

\textbf{Data Layer.} The data layer is widely involved in the database, transaction data, smart contract code, and other important information in the blockchain. It is responsible for the management of data storage and retrieval in the blockchain system. The data layer also includes the processing of encryption, compression, and indexing of data, with data storage and retrieval functions to ensure the security and integrity of data~\cite{mihoub2022denial}.

\textbf{Network Layer.} The network layer encompasses the network topology, routing protocols, peer-to-peer communication protocols, and other related components. The primary function is to facilitate the establishment and maintenance of connections and communications between nodes~\cite{katib2023blockchain}. This plays a pivotal role in determining the scalability, security, and performance of the blockchain network. The system must provide a secure and reliable communication environment to ensure that data can be exchanged and consensus reached between nodes.

\textbf{Consensus Layer.} The consensus layer is the core content of the blockchain system, including the consensus algorithm, the rules of the block generation, and the verification mechanism, providing the technical infrastructure for the blockchain~\cite{chithanuru2023anomaly}. The design of the consensus layer affects the security, performance, and decentralization of the blockchain system. It defines the rules and algorithms for reaching consensus among nodes and ensures that all nodes agree on the state of the blockchain.

\textbf{Incentive Layer.} The incentive layer includes incentives and reward rules that are designed to facilitate desired behaviors. Common incentive mechanisms include block rewards and transaction fees. Through the design of the incentive layer, it can promote the behavior of nodes in line with the interests of the system and guarantee the security and stability of the blockchain network~\cite{el2024ddos}.

\textbf{Contract Layer.} The contract layer includes the execution environment and related tools for smart contracts. It is mainly responsible for processing and executing the code of smart contracts. The existence of the contract layer provides a secure and trustworthy execution environment and ensures the correct execution and non-tamperability of smart contracts~\cite{liu2024repeated}.

\textbf{Application Layer.} The application layer is the interface for users to interact with the blockchain system, including the user interface, applications, and smart contracts. Users can interact with the blockchain system through applications, sending transactions, querying data, and other operations. The application layer provides users with a way to access the blockchain network, enabling blockchain technology to be applied in various scenarios, such as cryptocurrency trading and supply chain management~\cite{jia2025caa}.

\subsection{Common Security Threats}
Despite the numerous security advantages offered by blockchain technology, including decentralization, immutability, and transparency, there are still some common security threats to be considered \cite{duan2022multiple}.

\textbf{Double Spending.} Double spending is defined as the transmission of the same digital currency to two distinct addresses in a single transaction. This occurrence is typically attributable to network delays or exploitation of system vulnerabilities by attackers~\cite{parvathy2022certain}. This phenomenon may lead to a reduction in system trust, thereby eroding user confidence and, consequently, exerting an adverse effect on the overall reliability and trustworthiness of the blockchain system.

\textbf{51\% Attack.} 51\% attack is characterized by the attacker's possession of more than half of the computing power in the network, thereby enabling them to exert control over the consensus process within the network. The repercussions of these actions can encompass a range of issues, including the manipulation of transactions, the occurrence of double spending, and the disruption of service. In particularly grave instances, the integrity and stability of the entire blockchain network may be compromised~\cite{vats2024unveiling}.

\textbf{Smart Contract Vulnerability.} Smart contracts are based on code execution, and the existence of coding errors or vulnerabilities may lead to asset theft, contract deadlock, or abnormal execution. The ramifications of these actions include the forfeiture of assets, the execution of contracts that deviate from standard practices, and the compromise of system security. These factors exert a direct influence on the stability of the blockchain system and the confidence of its users~\cite{habib2023technique}.

\textbf{Loss or leakage of the private key.} The private key is how the digital asset is controlled. If the private key is lost or disclosed, attackers may gain control of the asset, which may result in theft of the asset, impersonation of the user's identity, and loss of control of the asset by the legitimate user~\cite{riadi2022optimization}.

\textbf{Leakage of transaction information.} Transaction information on blockchain is usually public, but sometimes users want to protect their privacy. After the transaction information is leaked, attackers can track the user's identity and behavior through their analysis of the transaction information, which leads to the exposure of the user's privacy, identity theft, and tracking of transaction behavior, and then violates the user's rights and interests~\cite{vats2024unveiling}.

\textbf{Fork Attack.} Fork attack refers to the occurrence of a fork in the blockchain network, which is used by attackers to implement malicious behaviors such as double-spending and canceling transactions. It may lead to chaotic transactions, compromised asset security, and reduced system credibility, directly threatening the stability and reliability of the blockchain system~\cite{parvathy2022certain}.

\textbf{Malicious Smart Contracts.} Malicious smart contracts are contracts designed to attack users, which may include theft of assets, denial of service, market manipulation, and other malicious behaviors, which can lead to loss of assets, damage to users' rights and interests, and decline in system trust, directly affecting the security of the blockchain system and the interests of users~\cite{jia2025caa}.

\subsection{DoS Attack}
Denial of Service (DoS) attack is a kind of by transmitting a large number of illegal application packets to the designated destination host, to occupy or consume the resources of the target host, so that the computer or network cannot provide normal services and the system of the target of the attack stops responding or even crashes \cite{raikwar2021attacks}. DoS attacks can be traced back to the early days of the Internet. As technology evolves, attackers continue to innovate and improve DoS attack methods to make them more threatening and complex, such as the SYN flood attack, the reflection amplification attack, and the DDoS attack.

DoS attack is a relatively low-cost form of cyberattack with the potential to have more severe consequences~\cite{duan2022multiple}. Attackers can use relatively simple tools and techniques to cause network performance degradation, node unavailability, smart contract unavailability, and data loss in blockchain systems.

\subsection{Classification of Common DoS Attack}
\subsubsection{SYN Flood Attack}
SYN flood attack is a common DoS attack, in which the attacker exploits a vulnerability in the TCP protocol's three-handshake process by sending a large number of forged SYN packets to the target server, but does not send subsequent ACK packets~\cite{ilyas2023prevention}. Since the server receives a large number of connection requests but is unable to complete the three handshakes, it will remain connected while waiting for the client's ACK packet. The attacker keeps sending forged connection requests, which makes a large number of half-connected states (SYN\_RCVD states) appear on the server, consuming server resources and ultimately causing the server to be unable to respond to normal user requests \cite{guru2023survey}. 

The following are common defenses:
\begin{itemize}
\item \textbf{SYN cookie.} It is a server-side application technology that generates a temporary cookie when establishing a TCP connection to replace the traditional half-open connection state, thus reducing the load on the server.
\item \textbf{Limit the number of connections.} Set firewall rules or server parameters to limit the number of connections to a single IP address to prevent too many connections from a single IP~\cite{yaish2024speculative}.
\item \textbf{Traffic filtering.} Filter suspicious SYN packets through network devices or firewalls to reduce the load on the server.
\item \textbf{Network device optimization.} Adjust network device parameters to increase network capacity and performance to handle a large number of connection requests.
\item \textbf{Monitoring and response.} Regularly monitor the status of the server to detect anomalies and take appropriate countermeasures.
\end{itemize}

\subsubsection{ICMP Attack}
The ICMP protocol is used to send diagnostic and error messages, while ICMP attacks utilize some features or vulnerabilities of the ICMP protocol to attack the target system, including destination unreachable, timeout, etc~\cite{aljanabi2022detect}. The attacker can use these messages to send a large number of false ICMP messages, thus consuming the network bandwidth and processing power of the target system and causing network congestion or resource exhaustion of the target system \cite{li2024guardians}.

The following are common defenses:
\begin{itemize}
\item \textbf{Firewall settings.} Set firewall rules to restrict access to the ICMP protocol.
\item \textbf{Traffic filtering.} Use network devices or firewalls to filter malicious ICMP messages.
\item \textbf{Enable ICMP Echo request response limit.} Set a limit on the target system to restrict the frequency or number of system responses to ping requests.
\item \textbf{Updates and patches.} Update the software of the system and network devices in a timely manner to fix known vulnerabilities in the ICMP protocol.
\item \textbf{Traffic monitoring.} Regularly monitor network traffic and system performance to detect anomalies and take the corresponding countermeasures promptly.
\end{itemize}

\subsubsection{Reflection Amplification Attack}
Reflection Amplification Attack (RAA) is one of the common DDoS attack methods that exploits the reflection properties and amplification effects in some network protocols, often utilizing protocols such as DNS, NTP, SSDP, and others. The attacker sends forged requests to open servers in the network, causing the servers to send a large number of responses to the victim, thus exhausting the network bandwidth \cite{chen2018system}.

The following are common defenses:
\begin{itemize}
\item \textbf{Filtering and blocking attack traffic.} Filter out traffic characterized by reflection amplification attacks through network devices or firewalls, and prevent attack traffic from entering the victim's network promptly.
\item \textbf{Restrict access to open server ports.} Implement access control lists (ACLs) or network access control (NACs) for publicly available server ports to restrict requests from unknown sources.
\item \textbf{Update and configure servers.} Update and configure servers regularly to patch known vulnerabilities and weaknesses and reduce the attack surface~\cite{dai2022ddos}.
\item \textbf{Network monitoring and traffic analysis.} Use network monitoring tools to monitor network traffic in real time, quickly detect anomalies, and take appropriate countermeasures.
\item \textbf{Protocol optimization and configuration adjustment.} Adjust network protocols and server configurations to limit or disable protocol functions that have amplification effects and reduce the impact of reflection amplification attacks.
\end{itemize}

\subsubsection{DDoS Attack}
DDoS attack refers to an attack in which an attacker controls a large number of infected computers, IoT devices, or servers located in different geographical locations, forming a massive attack network. These infected devices are referred to as "zombies" or "bots" \cite{li2024detecting}. The attacker remotely controls these devices, concentrating them to launch attacks, and uniformly sends a large amount of malicious traffic to the target system, rendering it unable to provide normal services. This attack method enables larger-scale attacks that are more difficult to trace and defend against \cite{zou2025malicious}.

The following are common defenses:
\begin{itemize}
\item \textbf{DDoS firewall and intrusion detection system.} Deploy specialized DDoS firewall and intrusion detection system to detect and block DDoS attack traffic in time.
\item \textbf{Traffic filtering and cleaning.} Use traffic filtering and cleaning services to filter out malicious DDoS attack traffic and deliver normal traffic to the target server.
\item \textbf{Load balancer.} Use a load balancer to spread the traffic to increase the capacity of the system and its ability to withstand DDoS attacks.
\item \textbf{DDoS attack monitoring and warning.} Regularly monitor network traffic and system performance to detect abnormal traffic and attacks in time and take the corresponding countermeasures.
\item \textbf{Update and maintenance.} Regularly update systems and applications to patch known vulnerabilities and weaknesses to reduce the risk of DDoS attacks.
\end{itemize}

\section{DoS Attacks in Blockchain Systems}
The blockchain network's decentralized architecture is characterized by the uniformity of status and role among all nodes. It is challenging to terminate the service of this blockchain network unless an attacker launches a coordinated attack on all nodes. If some nodes are omitted, the blockchain network is still able to operate normally \cite{singh2021blockchain}.

There are two reasons why traditional DoS attacks cannot be employed against blockchain networks~\cite{kumar2022distributed}. Firstly, nodes are dynamic and can access new nodes at any time, making it difficult for an attacker to find all nodes. Secondly, the presence of tens of thousands of nodes complicates the ability of an attacker to launch an attack on all of them simultaneously. Consequently, achieving a complete denial of service in blockchain systems is difficult. Instead, the objective of DoS attacks is to compromise or disrupt the functionality of the blockchain system.

\subsection{Network Layer} 
Denial-of-service attacks at the blockchain network layer aim to prevent the proper functioning of the blockchain network by taking up network bandwidth, disrupting communication between nodes, or rendering nodes inoperable, preventing communication between nodes and causing network fragmentation or delays~\cite{wang2022sdos}.

\subsubsection{Common Attack Principles}
\begin{itemize}
\item \textbf{Occupying bandwidth:} The attacker sends a large amount of network traffic to the target node or network, consuming bandwidth resources, resulting in network congestion or delay, so that legitimate users cannot be accessed normally.
\item \textbf{Blocking communication:} The attacker tries to prevent communication between nodes by sending error or malicious messages, or interfering with the network connection so that nodes cannot exchange data or participate in the consensus process.
\end{itemize}

\subsubsection{Distributed Denial of Service}
\begin{figure}
\centering
\includegraphics[width=0.68\linewidth]{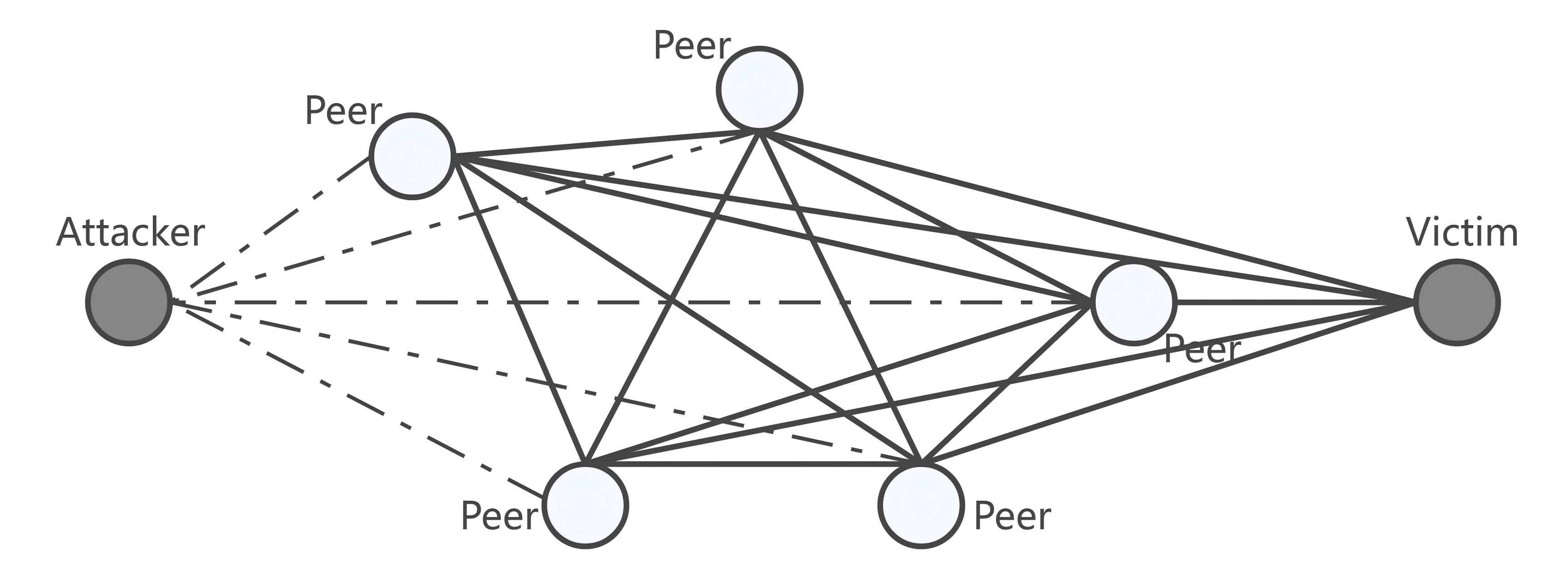}
\caption{\label{fig:DDoS Attack Process}DDoS Attack Process}
\end{figure}

Denial-of-service attacks typically necessitate the utilization of a substantial amount of network resources by the attacker to attack the target network or computer system. This results in the target system's inability to respond promptly to normal access, or even the system's failure~\cite{shah2022blockchain}. However, in the context of the blockchain system, the user nodes are tens of thousands, and the resources of each node are limited. Consequently, the attacker generally opts to integrate network bandwidth fragmentation to implement distributed denial-of-service (DDoS) attacks. The principle of DDoS attacks based on the blockchain network is shown in Figure \ref{fig:DDoS Attack Process}.

\subsection{Consensus Layer}
Denial-of-service attacks at the blockchain consensus layer aim to disrupt or interfere with the consensus mechanism of the blockchain network, thereby preventing nodes from agreeing on the correct blockchain state~\cite{chaganti2022comprehensive}. The most prevalent attack methods include the 51\% attack and the network splitting attack. In this section, we methodically analyze the 51\% attack as a paradigmatic case.

\subsubsection{51\% Attack Principle}
In blockchain, transactions and blocks are validated by numerous network nodes through a consensus process, thereby ensuring the integrity and validity of the transactions and blocks. The prevailing consensus mechanism is Proof of Work (PoW). In the PoW consensus protocol, the blockchain system allows the concurrent existence of multiple forked chains, with each chain having the capacity to externally assert its correctness \cite{ye2018analysis}.

However, if an attacker gains control of more than half (i.e., 51\%) of the network's computing power, they can significantly influence the network's behavior. To illustrate this phenomenon, one may envision a group of individuals engaged in a collective activity in which each participant is tasked with determining the subsequent action in unison. The collective decision-making process is facilitated by a voting system that ensures equitable representation for all participants. In the event that the majority agrees, the aforementioned action is to be executed. It is reasonable to hypothesize a scenario in which a single individual acquires a dominant share of the voting power in the game, which amounts to more than half of the total. This means that he can decide his next steps because he has enough votes to override the opinions of the others. He can choose to ignore the advice of the others altogether or make his own decisions that the others cannot oppose.

In blockchain, computing power is analogous to voting power. If an attacker controls more than 51\% of the computing power, he can control the network consensus process and agree on decisions and behaviors. This makes it possible for an attacker to carry out malicious behaviors such as double spending, blocking transaction confirmations, or tampering with transaction records, thus undermining the security and trustworthiness of the blockchain network.

\subsubsection{51\% Attack Process}
\begin{figure}
\centering
\includegraphics[width=0.85\linewidth]{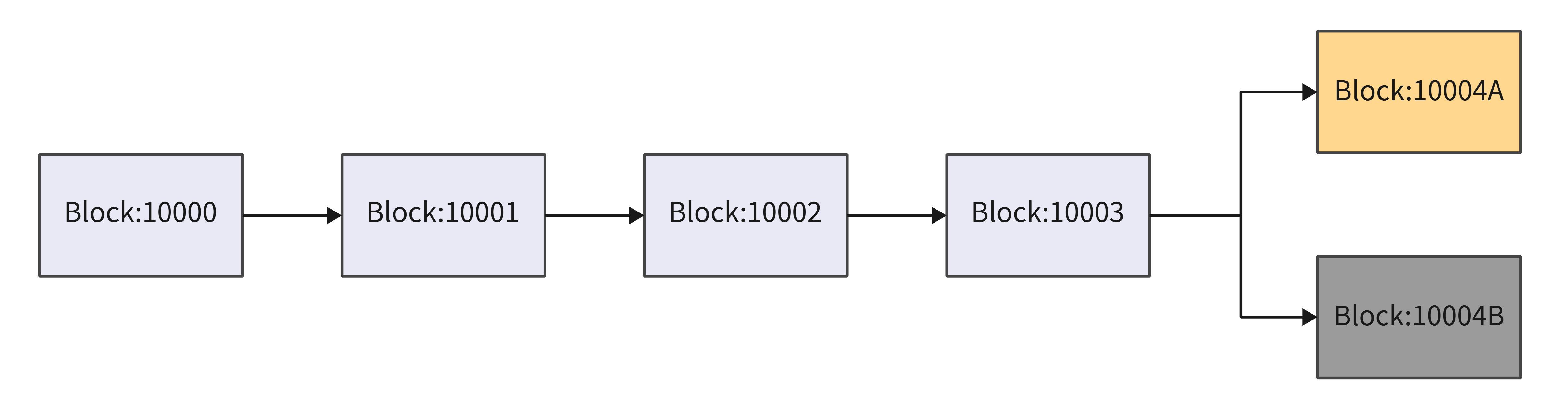}
\caption{\label{fig:Broadcast Fork}Broadcast Fork}
\end{figure}

It is hypothesized that an attacker, who exerts control over 50\% of the computing power, disseminates his initial transaction to half of the network and his subsequent transaction to the remaining half of the network. In each network segment, two miners almost simultaneously obtain accounting rights, and then the miners broadcast their respective accounting blocks. At this point, the initial unified ledger forks, yielding blocks A and B, as illustrated in Figure \ref{fig:Broadcast Fork}.

Subsequently, if the next miner chooses branch A to continue accounting, according to the PoW consensus mechanism, branch A, which is longer than branch B, will be recognized, while branch B will be discarded, as illustrated in Figure \ref{fig:Recognize Branch A, Discard Branch B}.
\begin{figure}
\centering
\includegraphics[width=0.95\linewidth]{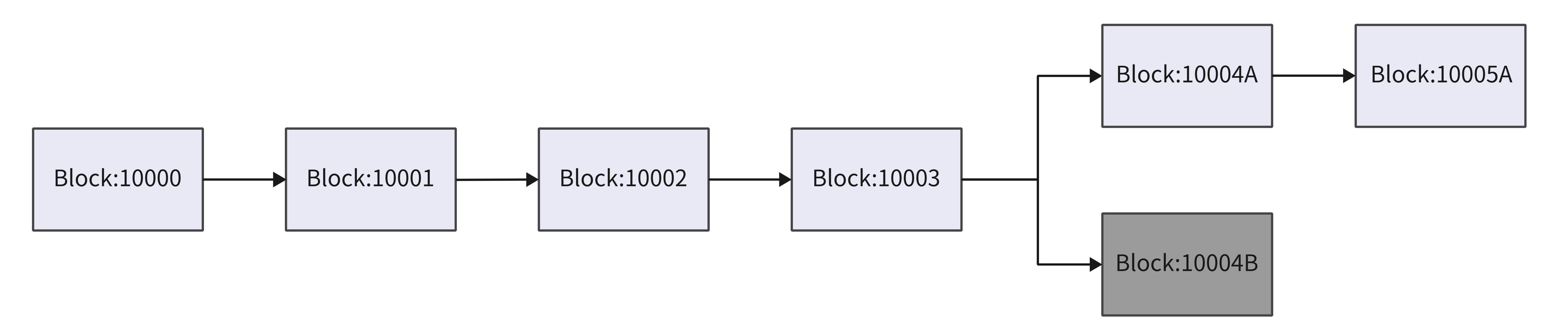}
\caption{\label{fig:Recognize Branch A, Discard Branch B}Recognize Branch A, Discard Branch B}
\end{figure}

\begin{figure}[t]
\centering
\includegraphics[width=0.99\linewidth]{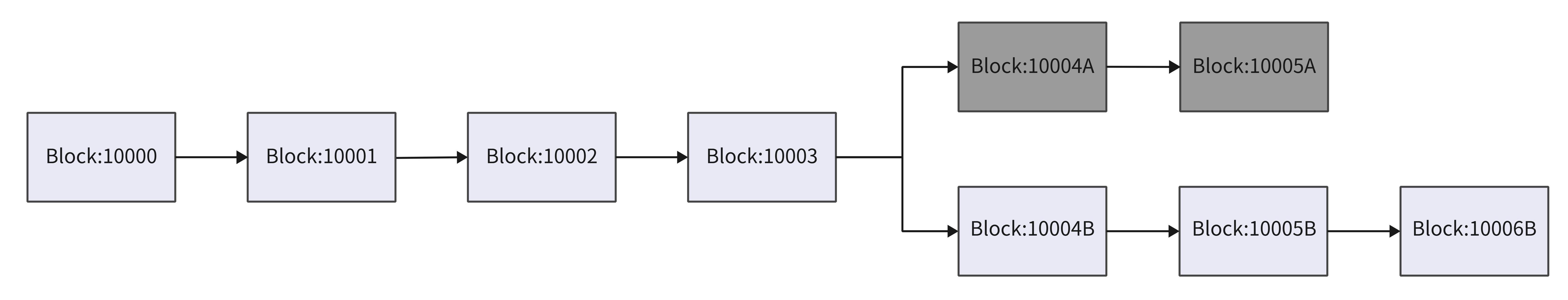}
\caption{\label{fig:Forced Renewal of Branch B}Forced Renewal of Branch B}
\end{figure}

If branch A is recognized, the first transaction is recognized. At this point, if the attacker gets the transaction item, he will use his arithmetic power to become a miner and perform two consecutive bookkeeping operations on the discarded branch B, as illustrated in Figure \ref{fig:Forced Renewal of Branch B}.

At this point, it is observed that branch B has a greater length than branch A. Consequently, branch A is then discarded, the transaction in branch A is not established, the first transaction recorded by the attacker in branch A is invalid, and the currency paid is returned to the original account. However, the attacker has obtained the goods. This means that he has expended one currency to obtain two goods, thereby successfully executing the double-spending attack.

\subsection{Data Layer}
Denial-of-service attacks at the blockchain data layer aim to make the network unable to provide services properly or create chaos by maliciously corrupting or tampering with blockchain data.

\subsubsection{Attack Principle}
Attackers perpetrate malicious operations on the data structures and transaction mechanisms of blockchain systems to consume network resources, interfere with the normal operation of nodes, or affect user operations. This results in nodes being prevented from processing data or verifying transactions normally, causing a denial of service or reduced system availability.

\subsubsection{Attack Methods}
\begin{itemize}
\item \textbf{Invalid transaction.} Attackers send a large number of invalid transactions, incorrectly formatted transactions, or transaction requests to the blockchain network. These transactions consume significant processing power and network bandwidth resources on the nodes, leading to network congestion or delays. These delays can further impede the nodes' ability to process valid transactions, resulting in delayed or discarded transactions.
\item \textbf{Tampering with data.} Attackers maliciously modify or tamper with data in the blockchain, including transaction records or status, transaction amounts, transaction initiators, transaction recipients, and other information, to interfere with or disrupt the system's transaction records and account balances. Attackers can modify the content of confirmed transactions or tamper with transaction information during the transaction broadcasting process, making it difficult for nodes to confirm the correct transaction records or leading to data inconsistency, which in turn affects the stability and security of the system.
\item \textbf{High-volume transaction attack.} Attackers send high-volume transactions, such as large money transfers or data transfer transactions, which makes nodes require more resources, such as computational and storage resources, to process these transactions. This results in a reduction in the processing speed and performance of the network, which affects its normal operation.
\end{itemize}

\subsection{Contract Layer}
Denial-of-service attacks at the blockchain contract layer aim to exploit vulnerabilities or design flaws in smart contracts, as well as limit performance during contract execution, to consume node resources, prevent transactions from taking place, or prevent the contract from functioning properly.

\subsubsection{Attack Principle}
An attacker exploits vulnerabilities or performance limitations of a smart contract to consume node resources or prevent normal interactions of other users for the purpose of consuming node resources, preventing transaction execution, or rendering the contract inoperable, thereby disrupting the functionality or services of the contract layer.

\subsubsection{Attack Methods}
\begin{itemize}
\item \textbf{Circular Call Attack.}
The attacker writes a malicious contract, which contains the operation of cyclic call or recursive call, calls the cyclic function therein, causing the contract execution to enter into a dead loop, which makes the node unable to end the contract execution normally, resulting in the node's resource exhaustion or crash.
\item \textbf{Resource exhaustion attack.}
The attacker writes a malicious contract that contains a large number of data storage operations, memory allocation, or computation operations, and uses resource-consuming operations or infinite loop operations in the contract to consume the node's storage space, memory, or computation resources, resulting in the node's inability to continue performing other operations.
\item \textbf{Denial of service attacks.}
An attacker writes a malicious contract that denies service by preventing other users from accessing or interacting with the contract normally by utilizing performance-limiting operations or error-handling mechanisms in the contract, such as refusing to accept a transaction, refusing to perform a function, or refusing to respond to a request.
\item \textbf{Contract status tampering.}
An attacker tampers with the state or data of a contract by sending a malicious transaction or calling a contract function that uses state variables or storage operations in the contract, resulting in inconsistent contract execution results or a contract that fails to operate properly.
\end{itemize}

\subsubsection{Smart Contract Attack}
A DoS attack on the smart contract means that an attack is used to stop the service, rather than a sustained network traffic attack. Attacking the smart contract requires continuous invocation, so direct attacks are infrequent. Instead, vulnerabilities in the smart contract code are typically exploited.

\textbf{(1) Attack Principle}

In many cases, attackers exploit vulnerabilities in network protocols to launch brute-force attacks, which can quickly deplete target resources. For example, if a website can handle a maximum of 100 user requests, an attacker could send 100,000 or more requests per second. In such cases, regardless of how fast the website server processes requests, how much memory it has, or how wide its network bandwidth is, the server will immediately crash or cease service. Even if the website does not crash, during the attack, normal users will be unable to access the website.

\textbf{(2) Attack Methods}

Denial-of-service attacks in smart contracts consist of at least six types: external calls without a set gas rate, calls relying on an external, owner misoperation, overly long arrays or mappings, logic design errors, and lack of dependent libraries.

\textbf{(3) Attack Reproduction}

In this section, a basic smart contract is used as an example to reproduce the attack. An auction contract is developed using the Solidity language in the Remix-Ethereum IDE. Subsequently, an attacker contract is designed to simulate the attack.

\textbf{A. Auction contract}
\begin{lstlisting}[language=Solidity]
contract Auction {
    address public winner;
    uint256 public amount;

    function bid() external payable{ 
        require(msg.value > 0,"amount is not 0");
        require(msg.value > amount,"amount is too small");

        payable(winner).transfer(amount);

        winner = msg.sender;
        amount = msg.value;
    }

    function balance() external view returns (unit256){ 
    }

    receive() external payable {}
}
\end{lstlisting}

\textbf{Code Analysis:}

The principle of bidding for contracts is that the highest bidder wins (i.e., WINNER). Consequently, the individual or entity with the highest bid is awarded the contract.

The address of the WINNER is:
\begin{lstlisting}[language=Solidity]
address public winner
\end{lstlisting}

The bid amount of the WINNER is:
\begin{lstlisting}[language=Solidity]
uint256 public amount
\end{lstlisting}

The bid function is modified with the payable keyword, indicating that the function can receive ETH (Ethereum). This means that when users call this function, they must also specify the bid amount.

\begin{lstlisting}[language=Solidity]
require(msg.value > 0,"amount is not 0");
\end{lstlisting}

This line of code constructs a conditional statement that requires the bid amount to be greater than 0.

\begin{lstlisting}[language=Solidity]
require(msg.value > amount,"amount is too small");
\end{lstlisting}

This line of code constructs a conditional statement that requires the bid amount to be greater than the highest bid amount currently recorded in the contract.

If the bid price is lower than the existing bid amount, the bid is rejected as it does not meet the requirements. If the bid price is higher than the existing bid amount, the program continues.

\begin{lstlisting}[language=Solidity]
payable(winner).transfer(amount);
\end{lstlisting}

A new winner is generated, and the transfer function is called to return the ETH of the previous winner.

\begin{lstlisting}[language=Solidity]
winner = msg.sender;
amount = msg.value;
\end{lstlisting}

Finally, the new winner and bid amount are updated.

\begin{lstlisting}[language=Solidity]
receive() external payable {}
\end{lstlisting}

The inclusion of the receive function indicates that the contract has the ability to receive tokens.

\textbf{Contract Loophole:}

The vulnerability in this contract is located mainly in the transfer function. If the refund is unsuccessful, the bid function will revert. At this time, the program is no longer running, with the entire function returning, and the bid is void. Accordingly, if there is a special address that makes the call to the transfer function fail, then the bidding function bid of the auction contract will become a pendulum and can not operate normally, resulting in the auction contract being scrapped. This is a DoS attack on the auction contract.

\textbf{B. Attacker contract}
\begin{lstlisting}[language=Solidity]
contract Attacker {
    constructor() payable{
    }
    
    function attack (address target,uint256 amount) external payable {
        Auction auction = Auction(payable(target));
        auction.bid{value;amount}();
    }
}
\end{lstlisting}

There are two types of Ethereum accounts: external accounts (EOA) and smart contract accounts. If a smart contract account contains a receive or fallback function, it can be used to make payments and receive payments. If not, it is restricted from receiving payments. Based on this, the attacker contract cannot be used to receive payments, while the auction contract can be used to receive payments.

\textbf{Code Analysis:}

\begin{lstlisting}[language=Solidity]
function attack (address target,uint256 amount) external payable {
\end{lstlisting}

This line of code constructs the attack function, where the parameter target represents the target address of the attack, i.e., the auction contract address, and the parameter amount represents the bid amount.

\begin{lstlisting}[language=Solidity]
Auction auction = Auction(payable(target));
auction.bid{value;amount}();
\end{lstlisting}

These two lines of code construct an auction contract object and call the auction contract's bid function to execute the bid.

\textbf{Attack Process:}

\textbf{1. Set the bid amount.} Set the bid amount to be slightly higher than the highest bid recorded in the current auction contract to bid successfully to become the winner. 

\textbf{2. Become the winner.} Once the attacker becomes the winner, others cannot bid. Assuming that the attacker is the first user to bid and the highest amount recorded in the current auction contract is 0, the attacker only needs to invest 0.01 ETH to immediately become the winner.

\textbf{3. Bid by others.} If there are other users to bid, the bid function is called, and the bid amount is 100 ETH. This bid amount is greater than 0 and greater than the current contract's highest bid amount of 0.01 ETH, so the contract continues. The execution of the following code will result in the repayment of the attacker's contract.
\begin{lstlisting}[language=Solidity]
payable(winner).transfer(amount);
\end{lstlisting}

However, since the attacker contract cannot be used to receive payments, the call to the transfer function fails, causing the entire function to stop running, roll back, and void the transaction. The attacker remains the winner, while others are unable to bid.

\section{Defense Mechanisms}
\subsection{Network Layer} 
The detection and resolution of denial-of-service attacks at the blockchain network layer mainly involve monitoring network traffic, identifying anomalous behaviors, improving network protocols, and implementing defensive strategies \cite{putz2020detecting}.

\subsubsection{Detection Methods}
\begin{itemize}
\item \textbf{Network traffic monitoring.} It can monitor incoming and outgoing data packets and detect abnormal traffic and large numbers of requests in the network. However, the cost of setting up and maintaining a monitoring system is high and requires professional technical support.
\item \textbf{Packet analysis.} It can analyze the header information and load content of network packets to identify abnormal packets and abnormal behaviors, such as a large number of invalid requests and high-frequency requests. However, the analysis process may consume a large amount of computing resources and time, affecting real-time performance.
\item \textbf{Connection status monitoring.} It can monitor the process of establishing and closing network connections, track the status and life cycle of connections, and quickly detect abnormal connections and abnormal connection behavior. However, monitoring connection status requires real-time collection and processing of large amounts of connection information, which may consume system resources.
\item \textbf{Inter-node communication monitoring.} It can monitor the communication process between nodes and detect abnormal nodes and abnormal messages, including node discovery, handshake process, and message passing. However, analyzing and processing data requires a certain amount of technical and human resources.
\end{itemize}

\subsubsection{Defense Methods}
\begin{itemize}
\item \textbf{Packet filtering and limiting.} It can filter and restrict abnormal data packets, intercept malicious requests or invalid data packets, prevent them from entering the network, and reduce the impact on nodes. However, it cannot address advanced attacks such as distributed denial-of-service attacks.
\item \textbf{Connection management and denial-of-service defense.} It can limit the number or frequency of connections from a single IP address, preventing malicious nodes from consuming network resources through large numbers of connections. However, it may mistakenly intercept normal user requests, affecting the user experience.
\item \textbf{Reverse proxy and load balancing.} It can distribute network traffic and requests, improve system stability and availability, and respond to sudden traffic spikes and DDoS attacks. However, it may be limited in the face of large-scale attacks.
\item \textbf{Address filtering and blocking.} It can blacklist or block malicious IP addresses, restrict their access to the network, and prevent them from attacking the network. However, it cannot prevent attacks from attackers using proxies or dynamic IP addresses.
\item \textbf{Network protocol improvement.} It can enhance the resistance to attacks and the security of the network, serving as a long-term solution. However, it requires extensive research and testing, which may take a considerable amount of time to implement and could potentially impact existing network architectures and equipment.
\item \textbf{Real-time response and automated defense.} It can detect and respond to network attacks promptly, automatically intercepting malicious traffic and malicious behavior. However, false positives or false negatives may occur, affecting normal system operation.
\item \textbf{Node diversity and decentralization.} It can reduce the single point of failure and attack surfaces, increasing the network's resistance to attacks and resilience. However, it has certain requirements for network topology design and optimization.
\item \textbf{Key management and authentication.} It can effectively prevent malicious nodes from intruding and impersonating, thus enhancing network security and reliability. However, it requires the establishment of a robust key management and identity authentication system, which comes at a relatively high cost.
\end{itemize}

\subsection{Consensus Layer}
The detection and resolution of denial-of-service attacks at the blockchain consensus layer mainly involve monitoring the consensus process, identifying anomalous behaviors, improving the consensus algorithm, and implementing defensive strategies \cite{yadav2023comparative}.

\subsubsection{Detection Methods}
\begin{itemize}
\item \textbf{Consensus process monitoring.} It can monitor message transmission and communication between nodes during the consensus process, detect the execution status of the consensus algorithm and node behavior, and quickly identify consensus interruptions or abnormal situations. However, false positives or false negatives may occur, requiring careful adjustment of the monitoring strategy.
\item \textbf{Network traffic analysis.} It can analyze network traffic patterns and packet transmission rates to identify anomalous traffic and high-frequency requests, and discover attacks that may lead to consensus disruption. However, the analysis process may involve some delay and cannot respond to real-time attacks promptly.
\item \textbf{Node status monitoring.} It can monitor the operational status and performance indicators of the nodes, thus enabling the timely detection of abnormal nodes and instances of resource exhaustion. However, it is imperative that monitoring metrics and thresholds are updated promptly. Conversely, the absence of such a measure may result in the occurrence of false positives or false negatives.
\item \textbf{Message validation and auditing.} It can ensure the integrity and correctness of messages, identify messages that may have been tampered with or forged, and prevent malicious interference with the consensus process. However, the audit process may experience some delay and may not be able to detect real-time attacks promptly.
\end{itemize}

\subsubsection{Defense Methods}
\begin{itemize}
\item \textbf{Byzantine Fault Tolerance mechanism.} The deployment of Byzantine Fault Tolerance (BFT) algorithms or other strong consistency consensus algorithms can improve the system's fault tolerance against malicious attacks and abnormal behavior, ensuring the stability and security of consensus. However, the implementation process is relatively complex and requires a deep understanding and optimization of the algorithm.
\item \textbf{Randomness introduced.} Introducing random elements into the consensus process, such as randomly selecting validators and delaying rounds, can reduce the predictability of attackers and increase the cost of attacks. However, the mechanism for introducing randomness must be carefully designed to avoid introducing new security risks.
\item \textbf{Message filtering and blocking.} It can exclude malicious nodes or tampered messages, maintaining the purity and credibility of the consensus process. However, there are certain requirements for the design and optimization of filtering rules that need to take into account the issues of misjudgment and missed judgment.
\item \textbf{Consensus strategy update.} It is a flexible defense mechanism that can adapt to constantly changing security threats. However, it requires timely tracking of the latest security research and attack techniques, which may require a significant investment of human and material resources.
\end{itemize}

\subsection{Data Layer}
The detection and resolution of denial-of-service attacks at the blockchain data layer mainly involve identifying anomalous behavior, implementing defensive strategies, and improving system architecture.

\subsubsection{Detection Methods}
\begin{itemize}
\item \textbf{Transaction record analysis.} It can identify abnormal trading behavior, such as a large number of invalid transactions, high-frequency trading, and non-compliant transaction formats. However, some attacks may use concealment techniques, making it difficult to identify abnormal transactions.
\item \textbf{Smart contract audit.} It is a preventive measure that can prevent malicious operations from causing denial-of-service attacks on the data layer. However, the audit process may take a long time, affecting the system's launch or update speed.
\item \textbf{Abnormal behavior recognition.} It can identify abnormal nodes or patterns of user behavior through behavioral analysis technology, such as abnormal transaction frequency and high-frequency contract calls. However, behavioral analysis technology requires a certain amount of training and fine-tuning, and there may be false positives or false negatives.
\end{itemize}

\subsubsection{Defense Methods}
\begin{itemize}
\item \textbf{Traffic filtering and restriction.} It can effectively filter and restrict abnormal traffic, reducing the impact on nodes and improving network stability. However, it may misjudge legitimate traffic, causing legitimate users to be blocked.
\item \textbf{Smart contract upgrade.} It can quickly fix vulnerabilities and security issues, improving the security and stability of contracts.
\item \textbf{Gas cost control.} It can limit the consumption of resources during the execution of contracts, thus preventing the exhaustion of node resources by malicious contracts. However, gas fees must be set with precision as any inaccuracy may affect the normal execution of the contract.
\item \textbf{Node load balancing.} It can distribute network traffic and requests, improving the stability and effectiveness of the system, but requires high costs and technology.
\item \textbf{Abnormal behavior blocking.} It can implement automated systems to prevent abnormal behavior, such as automatically identifying and rejecting abnormal transactions or malicious contracts. However, it needs to be constantly updated to respond to new types of attacks.
\item \textbf{Security policy updates.} It can enhance the system's ability to detect and respond to new types of attacks, thus ensuring network security. However, it is necessary to continuously monitor new technologies and threats in the security field and update security policies and rules accordingly.
\end{itemize}

\subsection{Contract Layer}
The detection and resolution of denial-of-service attacks at the blockchain contract layer mainly involve monitoring smart contract execution, identifying abnormal behavior, improving contract design, and implementing defense strategies \cite{alshudukhi2023interoperable}.

\subsubsection{Detection Methods}
\begin{itemize}
\item \textbf{Contract execution monitoring.} It can directly monitor the execution process of smart contracts, including contract function calls and transaction processing, track the time and resource consumption of contract execution, and promptly detect abnormal behavior. However, monitoring the contract execution process is costly and has a certain impact on system performance.
\item \textbf{Contract call analysis.} It can analyze contract invocations and parameter passing, identify abnormal or frequently invoked contracts, and detect attack behaviors that may cause contract execution interruptions. Compared to directly monitoring the contract execution process, it has lower costs and complexity but may not be able to detect certain abnormal situations promptly.
\item \textbf{Gas cost monitoring.} It can detect contract operations or transactions with high gas fee consumption and identify attack behaviors that may lead to resource depletion. Gas cost monitoring is relatively simple, with low costs, but it only reflects the resource consumption of the contract and cannot provide a comprehensive understanding of the contract execution process.
\item \textbf{Contract event monitoring.} It can monitor the triggering and handling of contract events, track contract status changes and abnormal events, and promptly detect attack behaviors that may cause contract interruptions. However, it may not be able to promptly detect certain abnormal situations that do not trigger events.
\end{itemize}

\subsubsection{Defense Methods}
\begin{itemize}
\item \textbf{Gas cost control.} It can limit the resource consumption of contract execution and improve the stability and availability of contract execution. However, it may affect the functionality and performance of the contract, resulting in restrictions on some normal operations.
\item \textbf{Contract code audit.} It can detect and fix vulnerabilities and security issues in contract code, improving contract security and reliability. However, the audit process can be time-consuming, affecting the speed at which contracts are launched or updated.
\item \textbf{Abnormal behavior detection and blocking.} It can detect and prevent malicious contracts or transactions promptly, monitor contract execution in real time, and respond quickly to abnormal events. However, it requires a certain amount of system resources, which may increase the burden on the system.
\item \textbf{Dynamic gas cost adjustment.} Dynamically adjusting gas fees based on contract execution status and network load conditions ensures fairness and stability of contract execution and prevents malicious attacks on contracts. However, implementation is complex and requires consideration of the execution status of the contract and network load conditions.
\item \textbf{Contract upgrades and rollbacks.} It is an emergency response measure that can take swift action when an attack is detected to ensure the normal operation of contracts and data security. However, it must be operated with caution to avoid causing a greater impact on the system.
\item \textbf{Abnormal event handling.} It can suspend contract execution or restore normal functionality to prevent further losses. It requires manual intervention and a certain amount of experience and technical support.
\item \textbf{Key management and privilege control.} It can restrict contract access permissions and scope of operations, preventing malicious contract intrusion and abuse, thus improving contract security and stability. However, it may increase system complexity and affect contract flexibility and scalability.
\end{itemize}

\subsubsection{Contract Vulnerability Exploitation Process Analysis}
Smart contracts, which function similarly to applets that run on the blockchain to formulate and execute contracts, are distinguished by open code, small code size, and asset involvement compared to traditional applications \cite{niu2024unveiling}. Concurrently, the code of a smart contract is incorporated as a transaction into a block of the blockchain. Consequently, smart contracts bear a resemblance to ordinary money transfer transactions in that the code or source code is publicly available and can be viewed at any time and in any location. In contrast to conventional applications, smart contracts are exclusively responsible for executing specific business logic on the blockchain. They are required to be incorporated into a block, a process that incurs computing and data billing charges, which can be substantial. Consequently, other business logic that is not intrinsically linked to the blockchain is typically not incorporated into smart contracts, resulting in relatively compact code volumes.

The three characteristics of smart contracts are also the easiest targets for hackers. The code is public, which means that there is no need to engage in extensive cracking and reverse analysis. The concise code requires minimal energy expenditure, and the cracking cycle is brief. Furthermore, there are assets involved, which can be directly profited from.

In subsequent practice, the code used contains a timestamp vulnerability, which, while it cannot cause the blockchain system to cease functioning, can result in a runtime error if exploited by an attacker. This, in turn, can render the contract unavailable and prevent it from providing normal service, i.e., denial of service.

\textbf{(1) Preliminary Preparation}

Environment setup: Install Ubantu on Windows 10 system virtual machine VMware environment, install Docker and Mythril on Linux system.

\textbf{(2) Tool Preparation}

\begin{figure}[ht]
\centering
\includegraphics[width=0.8\linewidth]{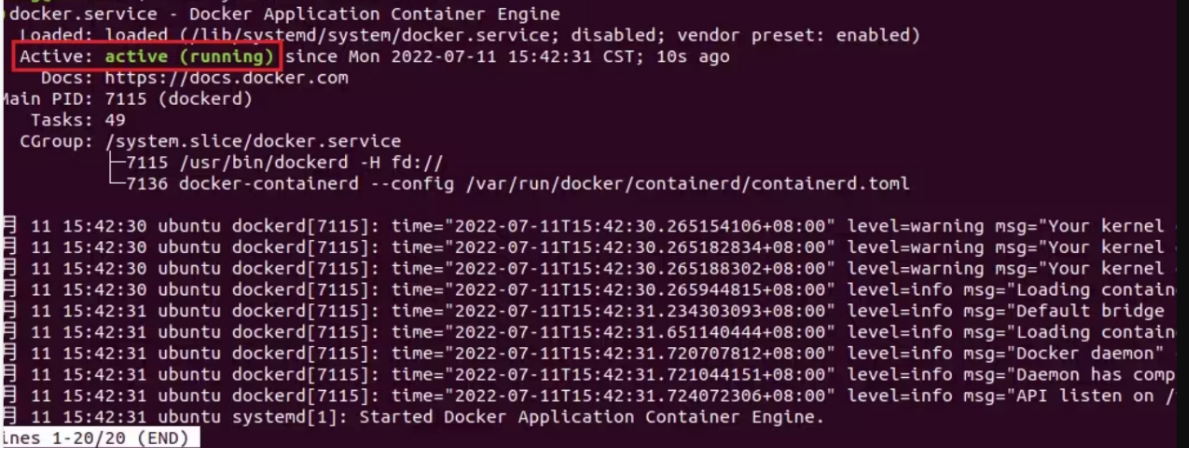}
\caption{\label{fig:Docker console output}Docker console output}
\end{figure}

Start the Docker service: the console output is shown in Figure \ref{fig:Docker console output}.

\begin{figure}
\centering
\includegraphics[width=0.8\linewidth]{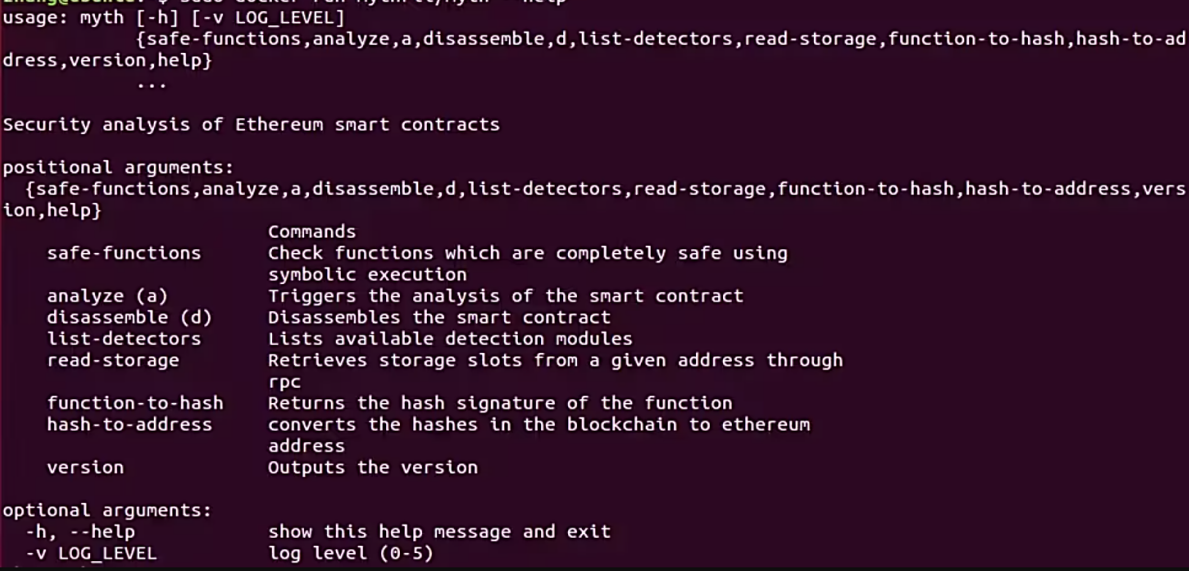}
\caption{\label{fig:Mythril output}Mythril output}
\end{figure}

Install Mythril and view the help: the console output is shown in Figure \ref{fig:Mythril output}.

\textbf{(3) Preventive Simulation}

The following is a smart contract code for simulation purposes. This code is similar to a lottery system with the following contract logic: players send the corresponding number of tokens (1 Ether in this contract) to the contract, which then performs a logical check. It takes the remainder of the blockchain timestamp modulo 15 within the same block. If the result is 0, the contract sends the remaining Ether in the contract as a reward to the player.
\begin{lstlisting}[language=Solidity]
// SPDX-License-Identifier: MIT
pragma solidity >=0.4.22;

contract Roulette {
    uint public pastBlockTime;

    // initially contract
    constructor() {}

    // receive function
    receive() external payable {}

    // fallback function used to make a bet
    fallback() external payable {
        require(msg.value == 1 ether); //must send 1 ether to play
        require(block.timestamp != pastBlockTime); //only 1 transaction per block
        pastBlockTime - block.timestamp;
        if(block.timestamp % 15 == 0) { // winner
            payable(msg.sender).transfer(address(this).balance);
        }
    }
}
\end{lstlisting}

\textbf{(4) Detection and analysis}

\begin{figure}
\centering
\includegraphics[width=0.68\linewidth]{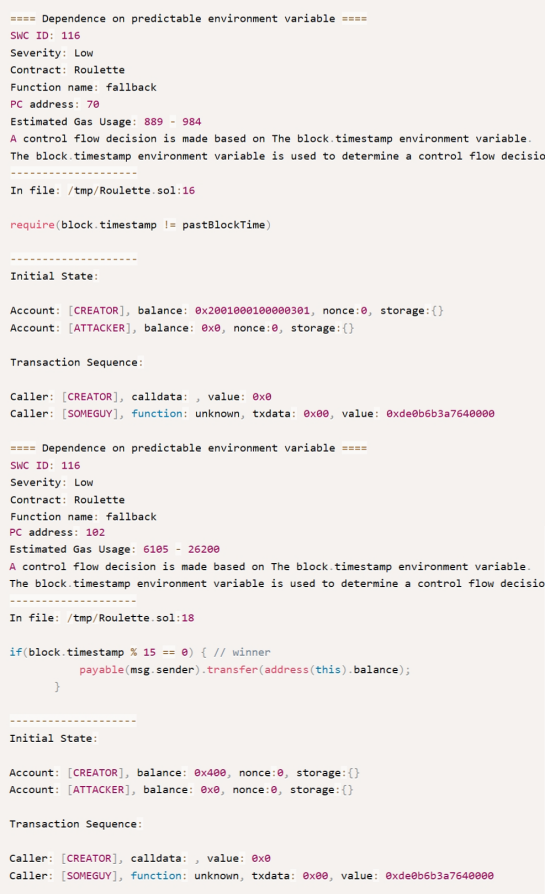}
\caption{\label{fig:Detection results}Detection results}
\end{figure}

Run the following command to analyze smart contracts using Mythril for detection:
\begin{lstlisting}[language=Solidity]
$ docker run -v $(pwd):/tmp mythril/myth analyze /tmp/Roulette.sol
\end{lstlisting}

The detection results are shown in Figure \ref{fig:Detection results}.
Based on the results of the detection and analysis, Mythril indicated that the contract code contained two identical timestamp dependency security vulnerabilities.

\textbf{The First Security Vulnerability:}

Vulnerability name: Dependency on predictable environment variables (timestamp dependency)

\begin{figure}
\centering
\includegraphics[width=0.68\linewidth]{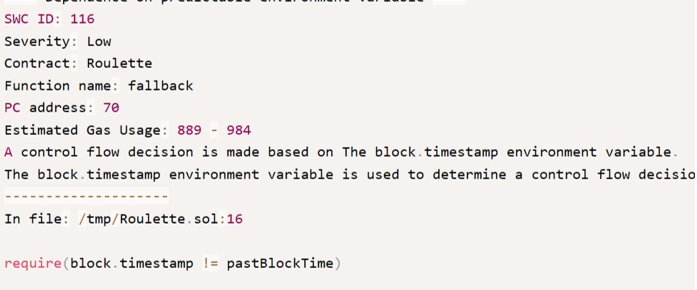}
\caption{\label{fig:Vulnerability 1}Vulnerability 1}
\end{figure}

The detection results of vulnerability 1 are shown in Figure \ref{fig:Vulnerability 1}.
From top to bottom, the information is displayed as follows: the vulnerability classification number, severity level, contract name, vulnerability function name, program counter address, estimated gas cost, and the line number and code segment of the vulnerability code in the contract code, as well as the contract's initial state and transaction sequence.

\textbf{The Second Security Vulnerability:}

Vulnerability name: Dependency on predictable environment variables (timestamp dependency)

\begin{figure}
\centering
\includegraphics[width=0.68\linewidth]{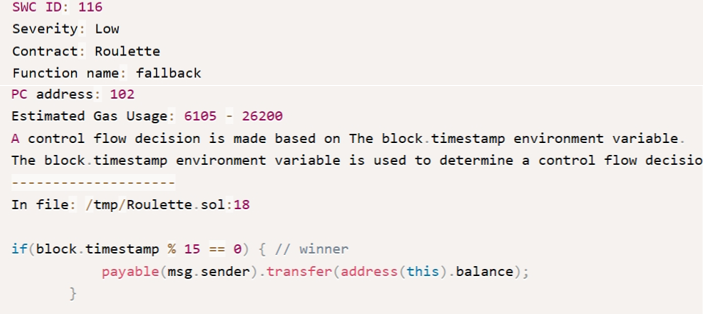}
\caption{\label{fig:Vulnerability 2}Vulnerability 2}
\end{figure}

The detection results of vulnerability 2 are shown in Figure \ref{fig:Vulnerability 2}.
From top to bottom, the information is displayed as follows: the vulnerability classification number, severity level, contract name, vulnerability function name, program counter address, estimated gas cost, and the line number and code segment of the vulnerability code in the contract code, as well as the contract's initial state and transaction sequence.

\section{Challenges and Limitations}
Blockchain technology plays an important role in changing traditional patterns, promoting digital transformation, strengthening data security, and achieving social justice. Due to its revolutionary technology and powerful advantages, it is widely and rapidly used in various industries, which undoubtedly brings many positive impacts and opportunities for the development of our society. However, we must not ignore that the blockchain system DoS attack and defense technology is still facing difficulties and challenges, such as \cite{chaganti2022comprehensive}:

\begin{itemize}
\item \textbf{Complex and varied means of attack.} DoS attacks have a variety of means, and attackers constantly change their strategies and techniques, making defense more difficult. Attacks at different levels require targeted defense strategies.
\item \textbf{The emergence of new types of attacks.} With the development of blockchain technology, new types of DoS attacks are constantly emerging, some of which may exploit the unique mechanisms of blockchain to launch attacks, making it difficult for existing defense technologies to respond.
\item \textbf{Difficulty in fixing contract vulnerabilities.} Smart contracts, as an important part of blockchain systems, are often targeted by attackers. However, fixing vulnerabilities in smart contracts is often complex and difficult, making them easy for attackers to exploit.
\item \textbf{Higher defense costs.} Implementing effective defenses requires significant investment in human, material, and financial resources. This may be a challenge for some small-scale blockchain projects.
\item \textbf{Lack of comprehensive research.} Current research on DoS attacks and defense techniques for blockchain systems is still fragmented and lacks systematic and comprehensive research. Lack of comprehensive understanding and analysis may lead to the neglect of new types of attacks.
\end{itemize}

In the face of these difficulties and challenges, the academic community and government departments need to work together to continuously strengthen research and cooperation, promote the advancement and innovation of DoS attack and defense technologies of blockchain systems, and improve the security and stability of blockchain systems. At the same time, it is also necessary for society and individuals to make efforts to comply with laws and regulations and contribute to the creation of a good network environment.

\section{Conclusion and Future Work}
This paper examines blockchain systems from a layered perspective, focusing on the network layer, the data layer, the consensus layer, and the contract layer. It provides an initial exploration of several typical denial-of-service (DoS) attack methods and defense techniques within blockchain systems, summarized as follows.

\begin{itemize}
\item \textbf{Diversified attack types.} DoS attacks are of various types, including contract layer attacks, consensus layer attacks, etc., which are complex and difficult to deal with.
\item \textbf{Diversity of defense techniques.} For different types of DoS attacks, researchers have proposed a variety of defense techniques, including traffic filtering and restriction, smart contract upgrading, gas cost control, node load balancing, etc.
\end{itemize}

However, there are still shortcomings in this paper. For example, the attack types are relatively single, and the summarized and compared defense methods are more effective for a single attack, while there is no in-depth discussion on composite attacks. The research on DoS attacks and defense technology of blockchain systems will still face challenges in the future, but it also has a broad development space and prospects.
Therefore, future research should focus on the following.
\begin{itemize}
\item \textbf{Comprehensive research.} More comprehensive and systematic research is needed to analyze different types of DoS attacks and their defense techniques in depth, to provide a more comprehensive guarantee for the security of the blockchain system.
\item \textbf{New technology applications.} New technology applications, such as artificial intelligence, machine learning, etc., can be explored to combine the characteristics of blockchain systems and propose more effective DoS attack defense techniques.
\item \textbf{Practical application validation.} The proposed defense technologies will be verified through practical application to assess their effectiveness and feasibility in the actual environment, providing technical support for practical application.
\end{itemize}

\bibliographystyle{plain}
\bibliography{bibliography}

\end{document}